\documentstyle[prl,amsfonts,epsfig,twocolumn,aps]{revtex}

\begin{document}
\wideabs{

\title{Capillary condensation in disordered porous materials:
hysteresis versus equilibrium behavior.}

\author{E. Kierlik\( ^{1} \), P. Monson\( ^{2} \), M. L. Rosinberg\( ^{1} \), L. Sarkisov\( ^{2} \), and
G. Tarjus\( ^{1} \)}

\address{ $^{1}$Laboratoire de Physique Th{\'e}orique des Liquides\cite{AAAuth}, Universit{\'e}
 Pierre et Marie Curie, 4 Place Jussieu,\\ 75252 Paris Cedex 05, France \\
$ ^{2}$Department of Chemical Engineering, University of Massachussetts, Amherst,
MA 01003, USA}

\maketitle

\begin{abstract}
We study the interplay between  hysteresis and equilibrium behavior in
capillary condensation of  fluids  in mesoporous disordered  materials
via a mean-field density functional theory of a disordered lattice-gas
model.   The  approach   reproduces  all   major  features    observed
experimentally.  We show that  the simple  van   der Waals picture  of
metastability fails  due to the appearance   of a complex free-energy
landscape with  a  large number  of metastable states.  In particular,
hysteresis can occur  both with and  without an underlying equilibrium
transition, thermodynamic  consistency  is  not  satisfied  along  the
hysteresis   loop,  and  out-of-equilibrium   phase  transitions   are
possible.
\end{abstract}
}

Capillary condensation of  a gas inside a mesoporous   material
refers to  the rapid change  to a liquid-like state  that  occurs at a
pressure (or chemical potential) lower  than the bulk saturation value
\cite{1}\cite{2}. This phenomenon  is   often thought of as  a
shifted gas-liquid transition.  Theoretical studies of fluids confined
in single pores  of  ideal geometry  have clarified the  mechanism for
such a shifted transition   and  introduced the concept  of  capillary
criticality  that describes the  fact  that the liquid-vapor  critical
point    in a  pore   occurs  at  a  lower   temperature  than  in the
bulk\cite{2}\cite{3}.   However,  the situation   in  real  mesoporous
materials, such as porous glasses and silica  gels, that consist of an
interconnected network of  winding  pores of varying shape,  curvature
and  size is not as clear.   With the possible  exception of fluids in
aerogels of  very high porosity  ($95-98 \%$)  for  which a {\it  bona
fide}  liquid-gas phase transition has been   reported with no sign of
hysteresis\cite{4}, there is no direct evidence of a first-order phase
transition characterized,  say, by a   jump in the adsorbed  amount of
fluid, nor of  true criticality signaled   by density fluctuations  on
very large length scales.  The typical experimental observation is the
presence of  a hysteresis loop in  the sorption isotherms, a loop that
is reproducible and vanishes at a  temperature lower than the critical
temperature  of  the  bulk   fluid\cite{1}\cite{2}\cite{5}.   The main
questions  raised by  this phenomenology are:  is there  a true  phase
transition associated   with   capillary condensation   in  disordered
materials?  What  is the connection  between  this  transition and the
observed hysteresis?  What are the sources of hysteretic behavior?  No
satisfactory  answers  have been  provided  so far.   Indeed, in these
systems    that     are  locally   heterogeneous,  but macroscopically
(statistically) homogeneous,  the statistical-mechanical methods  that
have proved very  efficient in dealing with bulk  fluids and fluids in
simple external fields, namely computer simulations, integral-equation
theory   and   density  functional    theory,  are plagued     by many
difficulties.

In this letter, we present a mean-field density functional theory of a
disordered lattice-gas  model.   This model, introduced  in a previous
work\cite{6},     incorporates    the    main  physical    ingredients
characterizing fluids   in disordered mesoporous   media: preferential
adsorption of one phase of the fluid,  connectivity of the pore space,
geometric and energetic disorder, exclusion effect  due to the matrix.
(Note that the model could  also be used  to describe the demixing  of
binary   mixtures in   mesoporous  materials\cite{2}.)  The    present
approach allows us to study  for the first  time the interplay between
out-of-equilibrium  (hysteresis) and equilibrium behavior   associated
with capillary condensation in a disordered  matrix.  We show that the
theory  captures  the main aspects of   the phenomenology of capillary
condensation in real    systems.   The combination   of  disorder  and
interconnectivity of the pore network  drastically alters the  picture
of    capillary   condensation   built    upon  the   independent-pore
model\cite{1}\cite{2}\cite{7} or the simple  van der Waals description
of metastability because it  generates a complex free-energy landscape
characterized by a large number of  metastable states.  In particular,
we show that (i)  hysteresis can occur {\it with}  or {\it without} an
underlying equilibrium phase transition, (ii) the disappearance of the
hysteresis loop  is  not associated with  capillary criticality, (iii)
thermodynamic consistency is not  satisfied along the hysteresis loop,
and (iv) collective  out-of-equilibrium phenomena such as  macroscopic
``avalanches'' are possible.

The model is described by the following Hamiltonian,

\begin{eqnarray}
\label{eq:1}
\cal{H}=&-& w_{ff}\sum _{<ij>}\tau _{i}\eta _{i}\tau _{j}\eta _{j}
-w_{mf}\sum _{<ij>}[ \tau _{i}\eta _{i}( 1-\eta _{j})\nonumber\\
&+&\tau _{j}\eta _{j}( 1-\eta _{i})] ,
\end{eqnarray}
where \( \tau _{i}=0,1 \) and \(  \left( 1-\eta _{i}\right) =0,1 \) denote
the  fluid and matrix occupancy  variables, respectively, and the sums
run  over distinct pairs of   nearest-neighbor (n.n.)  sites. Although
different   disordered   microstructures of     the matrix   could  be
considered\cite{8}, we choose  here the  simplest nontrivial  case, a
random matrix. The model   is thus specified   by two parameters,  the
average matrix density \( \rho _{m} \) that fixes the porosity (equal to
\(    1-\rho _{m}  \))  and   the ratio of    the matrix-fluid  over the
fluid-fluid  interactions, \( y=w_{mf}/w_{ff}  \), that determines the
``wettability'' .     In the following  we   only  consider attractive
matrix-fluid interactions\cite{9} and we set
\( \rho _{m}=0.25 \).  We use a $3$D bcc lattice with linear size
$L=8,12,16,24,48$ for  which we consider many different configurations
of the random matrix, from $1400$ for $L=8$ to  $5$ for $L=48$.  Most
of the  results  are illustrated here for   $L=48$ and  typical matrix
configurations, but when necessary,  e.g.,  to check the existence  of
phase  transitions, we have    performed a finite-size  analysis after
averaging over all matrix configurations.

The  mean-field  density  functional  theory  (or  equivalently for  a
lattice, the local mean-field theory) starts with the formulation of a
free-energy functional of the fluid  density field for a given  matrix
realization \(  \left\{  \eta _{i}\right\}   \):

\begin{eqnarray}
&F&(\{\rho_i\})=\frac{1}{\beta}                                \sum_i[\rho_i\ln
\rho_i+(\eta_i-\rho_i)\ln(\eta_i-\rho_i)]\nonumber\\  &-&w_{ff}\sum_{<ij>}\rho_i\rho_j
- w_{mf}\sum_{<ij>}[\rho_i(1-\eta_j)+\rho_j(1-\eta_i)] \ ,
\end{eqnarray}
where $\beta=1/(k_BT)$ and $\rho   _{i}=<\eta  _{i}\tau _{i}>$ is   the average
fluid density at site \emph{i}. (The overall fluid  density is then \(
\rho _{f}=\left(  1/N\right) \sum _{i}\rho _{i}  \), where $N=2 L^3$  is the
total number of sites.) For  a given   chemical
potential
\( \mu  \), minimization of $\Omega(\{\rho_i\})=F(\{\rho_i\})-\mu \sum_i \rho_i$ w. r.  t.
the \(  \rho  _{i}  \)'s  provides  the  grand
potential $\Omega$ of the adsorbed fluid.  The  corresponding equations for the
fluid density on each site are

\begin{equation}
\label{eq:3}
\rho _{i}=\frac{\eta _{i}}{1+\exp \left[-\beta\left[ \mu +\sum _{j/i}\left( w_{ff}\rho _{j}+w_{mf}\left( 1-\eta _{j}\right) \right) \right]\right] },
\end{equation}
where the sum  is over all n.n.  of  site \emph{i}.  The  above set of
nonlinear coupled   equations  has been solved  by  means  of a simple
iteration algorithm according to two different protocols: (1) to mimic
the experimental procedure, we follow continuously the solutions under
small variations of the chemical potential \( \mu \) (typically, $\Delta \mu
=10^{-3} $)  and (2)  to  search  more  exhaustively the  solutions of
Eq. (3) for a given \( \mu \), we repeat the iteration procedure with a
large number of initial conditions (typically, $10^2$) corresponding to
uniform  fillings of the lattice  at different overall fluid densities
(i.e., $\rho_{i}^{(0)}=\rho^{(0)}$).

We  show   in  Fig.  1    typical  isotherms obtained    by increasing
continuously  \(  \mu  \)  from  \(   -\infty \) (adsorption   branch)  and
decreasing continuously \(  \mu     \) from  \( +\infty     \)  (desorption
branch). They look qualitatively similar to the experimental isotherms
of  fluids     adsorbed    in    porous    glasses     and      silica
gels\cite{1}\cite{2}\cite{5},  with an   asymmetric  hysteresis   loop
(corresponding  to   the   H2  shape  in  the  IUPAC   classification)
characterized  by a steep  desortion  branch
 
\begin{figure}
\begin{center}
\epsfxsize= 200pt
\epsffile{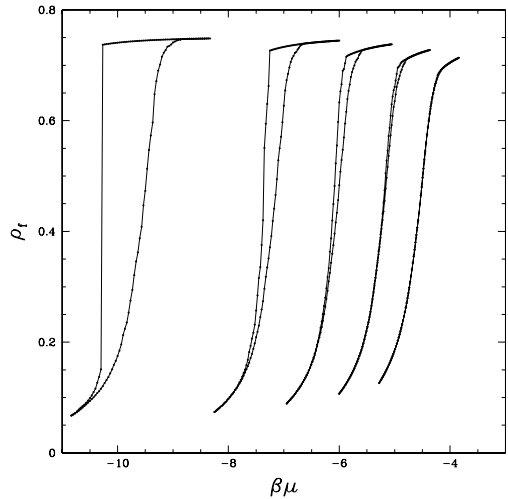}
\vspace{.4cm}
\caption{Theoretical sorption isotherms  for  $y=1.5  $  at  $T=0.6,0.8, 0.95, 1.1$   and
$1.25$. (The mean-field bulk critical point is at $T=2$.) }
\end{center}
\end{figure}

and  a    smoothly increasing    adsorption  branch.   The   so-called
``hysteresis phase diagrams'' that are constructed  from the points at
which the loop  opens and closes are  shifted to lower $T$ and  higher
$\rho_f$   compared to  the   bulk-fluid coexistence  curve,   and their
(asymmetric) shape is  also quite comparable  to what is obtained from
experimental data\cite{1}\cite{2}\cite{5}.

An  important test   for  the  relevance of  any   model  of capillary
condensation in  disordered  porous  materials is  the  shape of   the
so-called ``scanning curves'', i.    e., curves that are obtained   by
reversing  the sign of  the evolution of  \( \mu \) at different points
along the adsorption or  the desorption branches\cite{7}. It  has been
stressed that  these curves are not  properly reproduced by the widely
used  independent-pore model\cite{1}\cite{2}\cite{7}.  An illustration
of the scanning curves  on  adsorption and desorption  obtained within
the present theory is shown in Fig. 2. They look strikingly similar to
those observed experimentally, with an  upward curvature on adsorption
and  a downward curvature on   desorption.  When varying the  chemical
potential up and down  along different  paths we  find a  hierarchy of
inner scanning curves that display the ``return-point memory'' effect,
in close analogy to the hysteretic behavior displayed  by a variety of
systems\cite{10}\cite{11}\cite{12}.

The  above  results  confirm   that  the present approach   reproduces
qualitatively the phenomenology associated with capillary condensation
in porous glasses and silica   gels. This indicates that, despite  its
simplicity, the  model does   capture  the main physical  features  of
fluids adsorbed in real  mesoporous systems and that equilibration via
thermally   activated processes, processes     that are absent  in the
mean-field description, does not substantially modify the picture.
Indeed, the  implicit assumption behind
\begin{figure}
\begin{center}
\epsfxsize= 200pt
\epsffile{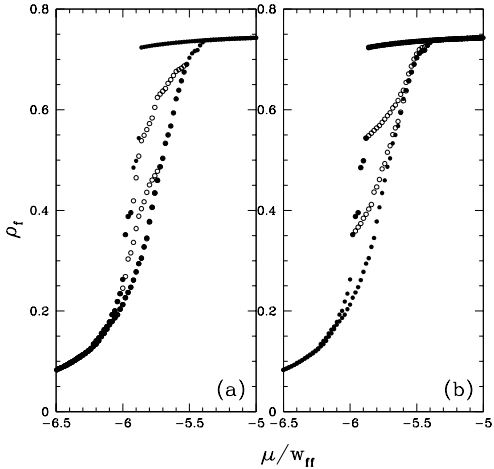}
\vspace{0.4cm}
\caption{Theoretical desorption (a) and adsorption (b) scanning curves
for \( y=1.5 \) and \( T=0.8 \).}
\end{center}
\end{figure}
the direct comparison
of the theoretical sorption curves with the  experimental ones is that
the system does not have time to equilibrate at constant
\( T \)  and \(  \mu \)  and  that adsorption/desorption only  proceeds
under  the influence of   chemical-potential changes. (This assumption
may break down when the perturbation induced  by the matrix is small as
in very dilute aerogels.)

Hysteresis    and   scanning     curves    are   manifestations     of
out-of-equilibrium behavior.  This is well  understood for instance in
systems like  athermal  martensites   and low-\(  T   \) ferromagnetic
materials whose transformations are driven by an external field; these
systems are well described  by  the \( T=0  \)  limit of models   with
quenched disorder\cite{10} (models somewhat related to the lattice-gas
studied  here).   On   the other  hand,  in  the   case  of  capillary
condensation  one often assumes that hysteresis  is also the signature
of a  true equilibrium phase transition  with two  possible metastable
states, gas and liquid, as in the standard van der Waals loop for bulk
fluids and fluids confined in a single pore.  However, this simple van
der Waals picture fails for the present system  because this latter has (in the
region where   capillary condensation   occurs) a complex  free-energy
landscape with a large  number of  metastable states,  as can  be seen
from  studying  the solutions of the   local mean-field equations, Eq.
(3).    For  each  $T, \mu$   and each  matrix   realization,   we have
investigated $50$ to  $100$ different initial  conditions to Eq.  (3).
When plotted  as in Fig.  3 on  the \(  \rho _{f}-\mu  \)  diagram (for a
typical matrix configuration of the $L=48$ system), all solutions fall
inside the major hysteresis  loop that is  found to coincide with  the
curves of extremal solutions obtained  from an initially empty lattice
(lower branch) and an initially filled lattice (upper branch). We suspect
 that this picture corresponds to the situation encountered 
in real  disordered porous materials (a similar behavior

\begin{figure}
\begin{center}
\epsfxsize= 200pt
\epsffile{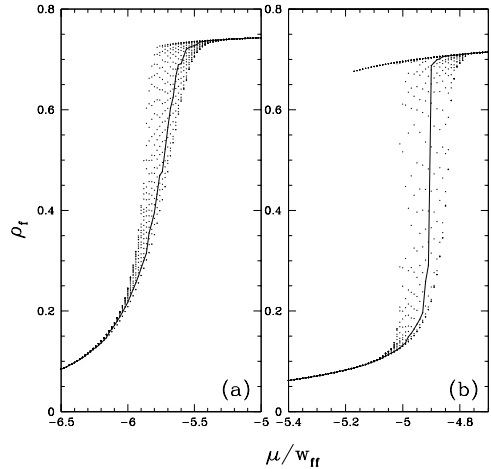}
\vspace{0.4cm}
\caption{Multiplicity of local mean-field solutions obtained from solving
Eq. (3) with many different initial conditions for each value of $\mu$:
(a) \( y=1.5 \) and \( T=0.8
\), (b) \( y=1  \)  and  \( T= 1\).    The solid lines represent   the
equilibrium curves obtained     by connecting the  states  of   lowest
grand potential.}
\end{center}
\end{figure}
has been observed for other disordered microstructures\cite{8}).

The most  significant  finding   of   our study is   that   hysteresis
associated with  capillary  condensation  in  disordered  porous media
occurs with  as well   as without   an underlying   equilibrium  phase
transition, the existence of this latter  depending on the strength of
the perturbation   induced  by  the matrix.    We  have   computed the
equilibrium   properties  of the   model  by  means  of  the  weighted
mean-field  approach\cite{13}:  at each $\mu$    and $T$, densities  of
extensive quantities  such as the fluid  density $\rho_f$ are calculated
as  Boltzmann  weighted sums   over  the different   solutions of  the
mean-field equations (i.e., each solution $\alpha$  is weighted by $e^{-\beta
\Omega^{\alpha}}/\sum_{\alpha}  e^{-\beta  \Omega^{\alpha}}$).  The  average   over the various
matrix realizations  is  taken afterwards.   For a  large matrix-fluid
interaction,   e.g.,   for $y=1.5$, there     is  no equilibrium phase
transition: this is illustrated    in Fig.  3a where  the  equilibrium
isotherm at   $T=0.8$  is  perfectly smooth  (it   stays so   at lower
temperatures).  On the other hand, for $y=0.5$, the system undergoes a
true  thermodynamic transition at low  $T$ as it becomes isomorphic to
the site-diluted  Ising model\cite{9} ($\rho_m=0.25$ is above the percolation threshold).  For
the intermediate value $y=1$, equilibrium  isotherms may display steep
portions, as illustrated in Fig.  3b for $T=1$ (however, a finite-size
scaling analysis   of  the results  for  $L=8,12$  and  $16$  seems to
indicate that this  isotherm is  still supercritical).  The  capillary
critical point certainly occurs at  a  \( T\) significantly below  the
temperature  \( T_{hyst} \)  at which hysteresis  first appears.  This
result is at odds with the behavior predicted  for a fluid confined in
a single pore\cite{2}\cite{3}  and is a   consequence of the  quenched
disorder; on the other hand, the fact  that true capillary criticality
may still   occur in disordered   porous  media is  at odds  with  the
prediction based on the independent-pore model\cite{1}\cite{2}\cite{7}
and results from the connectivity of the void  space accessible to the
adsorbed fluid.

\begin{figure}
\begin{center}
\epsfxsize= 240pt
\epsffile{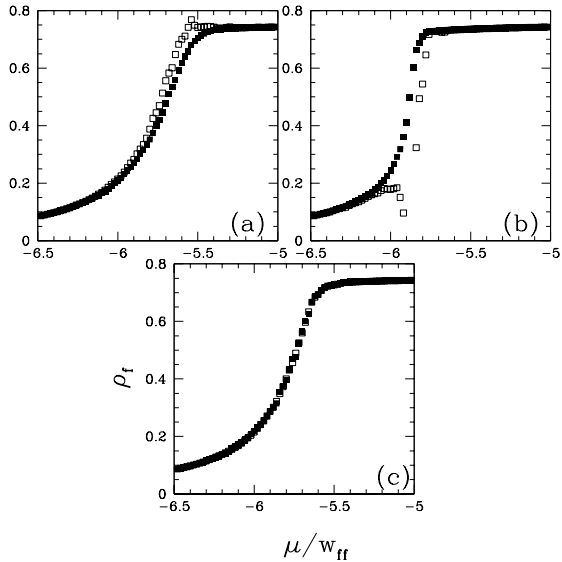}
\vspace{0.4cm}
\caption{Check of thermodynamic consistency (\( \left( \partial \Omega /\partial \mu
\right) _{T}=-N\rho _{f} \))\textbf{} along the adsorption (a), 
desorption    (b),    and   equilibrium    (c)  isotherms    ($y=1.5$,
$T=0.8$).  Filled symbols:  average  fluid density  obtained from  the
mean-field  solutions.      Open    symbols:  quantity    obtained  by
differentiating the corresponding grand potentials.}
\end{center}
\end{figure}

The breakdown of the simple van  der Waals picture of metastability in
the present problem shows up distinctly when considering thermodynamic
consistency along the sorption  isotherms.  Since the theory  provides
both the  grand  potential and the fluid   density, one can  study the
validity of the Gibbs adsorption equation,
\( \left( \partial \Omega /\partial \mu \right) _{T}=-N\rho_{f} \), along particular
isotherms.  As shown in  Fig.  4 a,b, the   equation is not satisfied,
neither  along the adsorption isotherm  nor  along the desorption one.
This  results from the  fact that the system  often jumps from a given
grand potential minimum to another along these  two isotherms and then
looses thermodynamic  consistency.  An  important consequence  is that
the thermodynamic integration  procedure used for building equilibrium
phase  diagrams\cite{2}   is no  longer   valid for  this  model.  For
instance,  in the case  \( y=1.5 \),  this latter procedure predicts a
capillary phase diagram whereas,  as already discussed, no equilibrium
phase transition takes place.  We believe this  may also be  true more
generally  for     other models  of   fluids   in   disordered  porous
materials. Note that the Gibbs adsorption equation is obeyed along the
equilibrium isotherm,  as illustrated in  Fig.  4c. (Interestingly, in
the cases   studied here, the   isotherm calculated from  the weighted
mean-field approach is  virtually indistinguishable from that obtained
by taking  at   each  $\mu$ the  solution  with   the  lowest  grand
potential.)

As a final  remark, we mention that  the mean-field density functional
theory   also predicts  the   occurrence of  out-of-equilibrium  phase
transitions.   Similarly  to the macroscopic   avalanches observed  in
low-\( T    \) ferromagnetic   materials\cite{10}, the   adsorption and
desorption branches of  the  major hysteresis loop may   display sharp
jumps. This is   illustrated in Fig.  1 where   one can see that   the
desorption  isotherm has  a   discontinuity at the  lowest temperature
shown. Due to the asymmetry of  the random chemical potentials exerted
by the matrix on the fluid, the jump always occurs at a higher \( T \)
in the desorption branch than in the adsorption branch. Whether or not
this out-of-equilibrium  transition,   with  its associated   critical
behavior\cite{10}, can be observed  in capillary condensation of fluids
in   real  mesoporous  materials depends  on   the  efficiency of  the
thermally activated processes.


\begin{references}
\bibitem{9} The system has an overall symmetry w.r.t. \( y=1/2  \):
this case corresponds to the site-diluted Ising model\cite{6}.
\end{references}
\end{document}